\documentclass{vak2024}
\usepackage[utf8]{inputenc}
\usepackage{url}
\usepackage{hyperref}

\begin{document}
\title{Star formation in low brightness galaxies and in the extended gaseous disks of normal galaxies}
\titlerunning{LSB-galaxies}  
\author{Anatoly~Zasov\inst{1}\and Natalia~Zaitseva\inst{1}\and Anna~Saburova\inst{1}}
\authorrunning{Zasov et al.} 
	%
	%
\institute{ Sternberg Astronomical Institute, Moscow State University, Universitetsky pr., 13, Moscow, 119234, Russia
	    }
\abstract{We analyze the available observational data on the radial distribution of gas and young stellar populations in the disks of low surface brightness (LSB) galaxies and the outer regions or the extended disks of normal brightness (HSB) galaxies. These  cases involve star formation under specific conditions of low volume and surface gas density. There is no well defined boundary between these subgroups of galaxies that we consider, but in non-dwarf LSB galaxies the rate of current star formation within the wide range of radial distances  appears to be higher compared to the outer disks of most of HSB galaxies at similar values of the surface gas density. The factors that could stimulate the compression of the rarefied gas at the periphery of galaxies are briefly discussed. Attention is drawn to the idea that the densities of LSB disks estimated from their brightness may be underestimated.

\doi{10.26119/VAK2024.041}
}

\maketitle

\section{Introduction}

There are two types of non-dwarf galaxies in which weak star formation can be traced to distances from the center where the surface gas density ($\Sigma_{HI}$) is very low and there are no obvious signs of ongoing interaction between the galaxy and its close surroundings that could provoke gas compression. First, there is a certain fraction of disky (usually spiral) galaxies with normal brightness (HSB-galaxies) (hereafter called normal galaxies) that have a continuation of gas disks beyond the optical radius $R_{25}$, and their faint UV emission suggests the presence of young stars (XUV disks, see Thilker et al. 2007). Second, they are low surface brightness galaxies (LSBs) with the extended disks of giant size, in which signs of star formation can be seen even at distances of tens of kpc from the center, despite the very shallow gravitational potential profile associated with the weak disk gravity. The formation of LSB galaxies and the extended disks of normal galaxies can involve several scenarios and is the
subject of active discussion (see e.g. Saburova et al. 2019; Saburova et al. 2021 and references in these papers). In this paper we compare the low density regions of HSB and LSB galaxies in terms of their stellar and gas densities and star formation rates.

\section{SFR vs stellar and gaseous densities of disks}

Star formation in the disks of LSB galaxies, as well as in the outer disks of some normal spirals, occurs under the specific conditions of a low density of the gas layer, which widens at the disk periphery, and a low angular velocity of disk rotation (the first factor hinders star formation, the second favors it). As in the disks of HSB galaxies, the star formation rate (SFR) in the disks of LSB galaxies correlates with the azimuthally averaged surface gas density $\Sigma_{HI}$, but for LSB galaxies this dependence is more blurred. Due to the low density of the disks, star formation there is not associated with the large-scale disk instability, but with the local sites of gas compression near the disk plane caused by other factors that perturb a gas layer.

\begin{figure*}
\centerline{\includegraphics[width=0.6\textwidth]{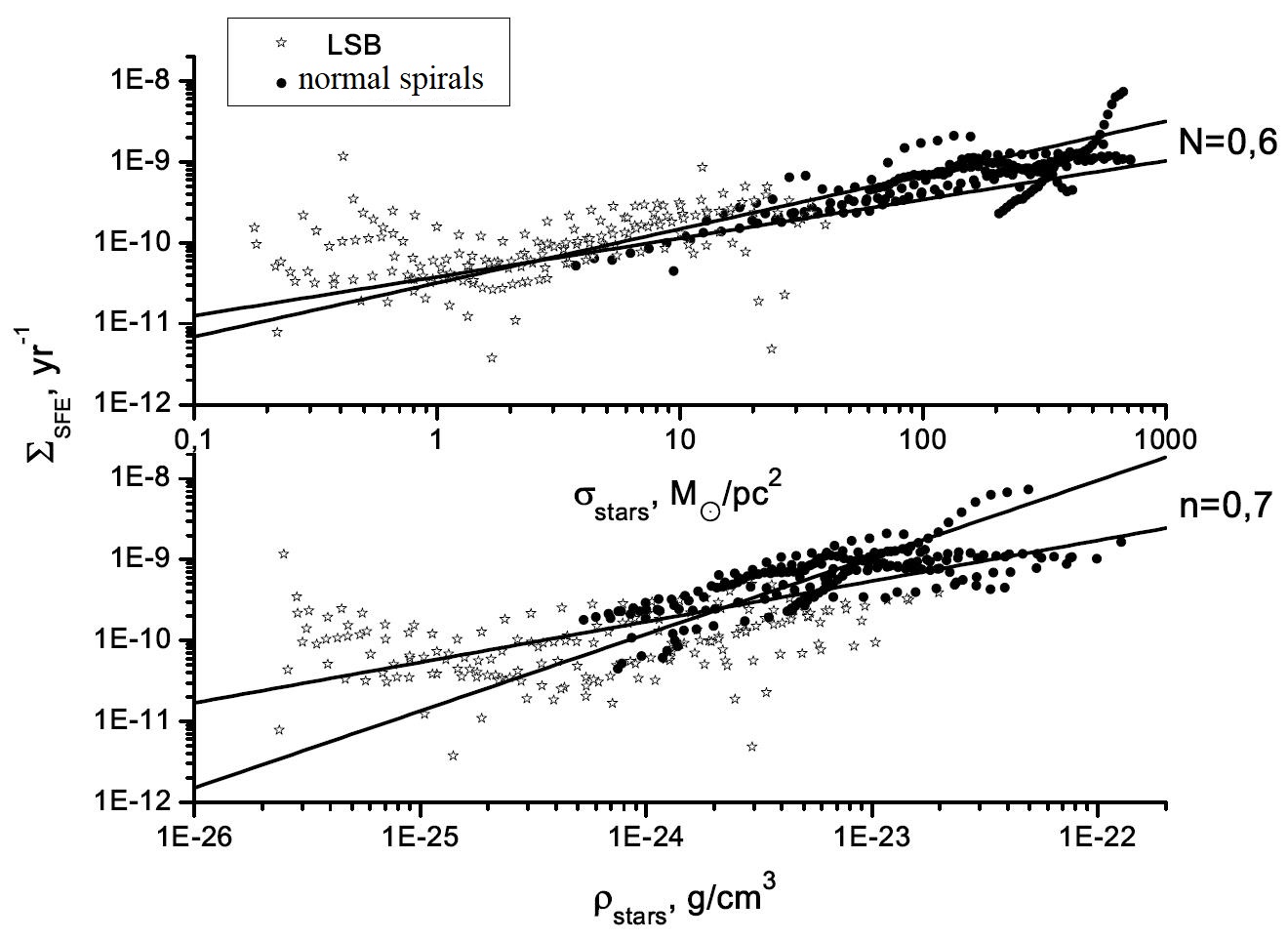}}
\caption{Correlation between $\Sigma_{SFE}$ and $\sigma_{stars}$ or $\rho_{stars}$ for two different sub-samples: LSB-galaxies (marked as grey pentargams) and normal spiral (marked as black dots). Straight lines correspond to regressions x(y) and y(x) applied to the set of all points by the method of least squares.}
\label{fig:1}
\end{figure*}

It is convenient to characterize the ability of the gas to give birth to stars by the value of the star formation efficiency $SFE = SFR/\Sigma_{HI}$, which represents a star formation rate per unit mass of HI per unit area of a disk. In general, the SFE decreases monotonically with radial distance R, so that its inverse, the gas exhaustion timescale, is very high, reaching tens of billions of years or more at the distant periphery even for HSB-galaxies. In Abramova $\&$ Zasov (2012) it was shown that the SFE at a given radius R in galaxies of different types is closely correlated not with the gas density, but with the surface and volume densities of a stellar disk in a disk plane where the normal and LSB galaxies form a single sequence (see Fig.\ref{fig:1}). The volume density calculations were performed for galaxies with known radial distributions of SFR, $\Sigma_{gas}$, and a rotation curve, assuming a fixed value of the gas velocity dispersion. They were reduced to a joint solution of the Poisson equation and the differential equations of hydrostatic equilibrium of a gas layer in the gravitational field created by disk components and a dark halo. At very low stellar disk densities ($\sigma_{stars} <3 M_*/pc^2$), the relation of SFR or SFE to the density of the stellar-gaseous disk becomes loose. The decrease of the SFE along the radius reflects the increasing thickness of a gas layer and consequently a decrease of the gas volume density, which is in good agreement with the model of Ostriker et al. (2010) of self-regulated star formation in an equilibrium layer of diffuse gas in the disk plane. However, the applicability of this model to extremely low-density disks is not obvious. There, star formation tends to be sporadic and weakly dependent on the disk properties.

It is worth comparing LSB-galaxies with the low density regions of normal brightness galaxies beyond the conventional optical radius $R_{25}$. The examples of relatively close galaxies with the weak extension of star forming disks are NGC 289, M83, NGC 628, NGC 2841, NGC 3521, NGC 5055, and some others. To the type of HSB- galaxies with high gas content and observed star formation in the extended disk one can also add the galaxies of the ``Bluedisk'' sample (Wang et al. 2013), and the ``HIX galaxy sample'' of gas-rich HSB-galaxies (Lutz et al. 2018). In fact, there is no sharp boundary between LSB galaxies and the outer low brightness disks of HSB galaxies. The example of a normal galaxy with a giant low-luminosity disk that allows us to refer it also to the LSB-type galaxies is UGC 1378, a massive SBa-type galaxy whose low brightness outer disk can be traced over 50 kpc, making it similar to giant LSB galaxies such as Malin-1, 2 (Saburova et al. 2019).

\begin{figure*}
\centerline{\includegraphics[width=0.55\textwidth]{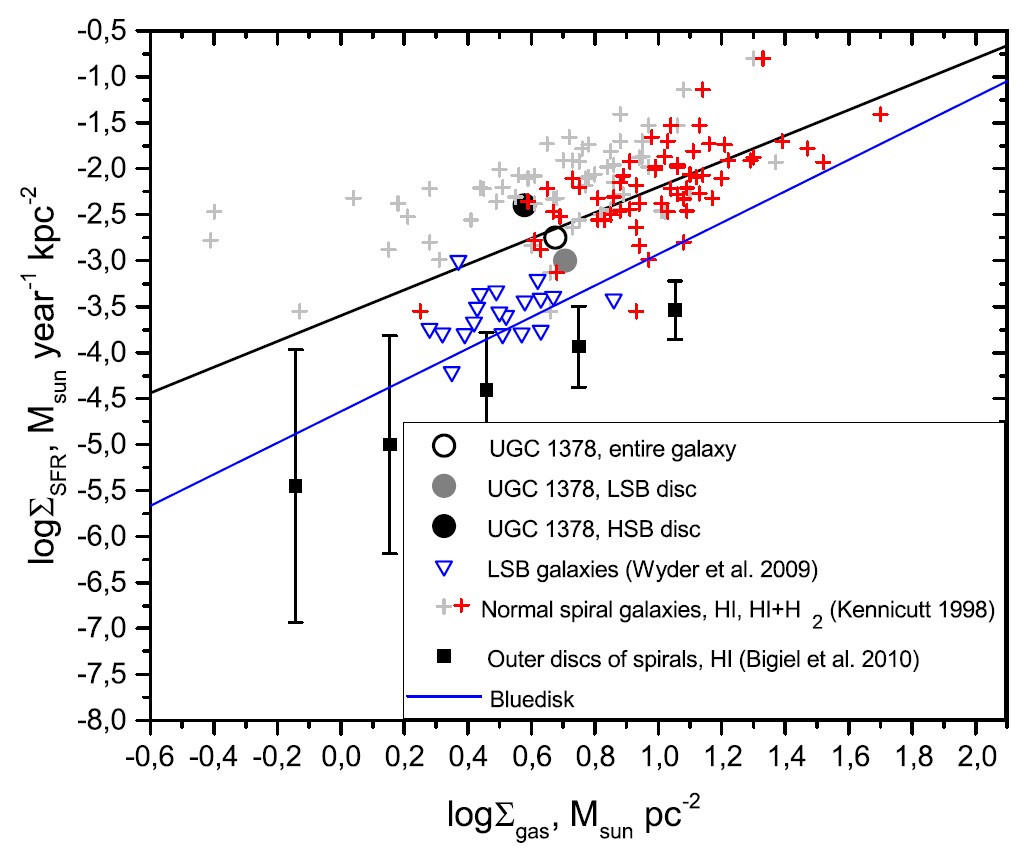}}
\caption{The SFR surface density versus gas surface density diagram. The sample of normal spirals is plotted as bright and faint crosses corresponding to total and HI gas densities, respectively. The black line corresponds to the Kennicutt-Schmidt law. The blue line is the fit for the Bluedisk galaxies. The circles mark UGC 1378. The empty triangles show the position of the LSB galaxies. The outer parts of normal HSB galaxies (black squares) have a lower SFR for a given gas surface density.}
\label{fig:2}
\end{figure*}

The mechanisms that drive the formation of star-forming regions in such rarefied disks are poorly understood. The rate of star formation there is determined there not only by the mean gas density, but also by how inhomogeneous the gas distribution is on the kiloparsec scale. In the most massive LSB-galaxies we observe a line of stellar spiral arms, but in other cases star formation takes place in randomly distributed areas of a disk which, if they are sufficiently large, can stretch into arcs due to differential rotation. If tidal interactions with nearby galaxies are excluded, strong local density inhomogeneities can be created by the accretion of gaseous flows or dwarf satellites, as well as by the passage of small sub-halos through a disk plane. The efficiency of star formation in these cases depends not only on the gas density, but mostly on the disk environment, which can be very different even for galaxies of the same type. 

The observed SFR of the LSB disks turns out to be rather high for their gas density. This is illustrated by Fig.\ref{fig:2} (Saburova et al. 2019),, which compares the averaged ($\Sigma_{SFR}$) and gas densities of normal (crosses) and LSB (empty triangles) galaxies, as well as the outer regions of normal spirals (black squares). As can be seen from the diagram, $\Sigma_{SFR}$ is lower in LSB galaxies than in HSB-galaxies, but tends to be higher for a given gas density compared to the outer disks of normal galaxies, most likely due to the lower disk thickness which makes  $\rho_{gas}$ higher.

\begin{figure*}
\centerline{\includegraphics[width=0.65\textwidth]{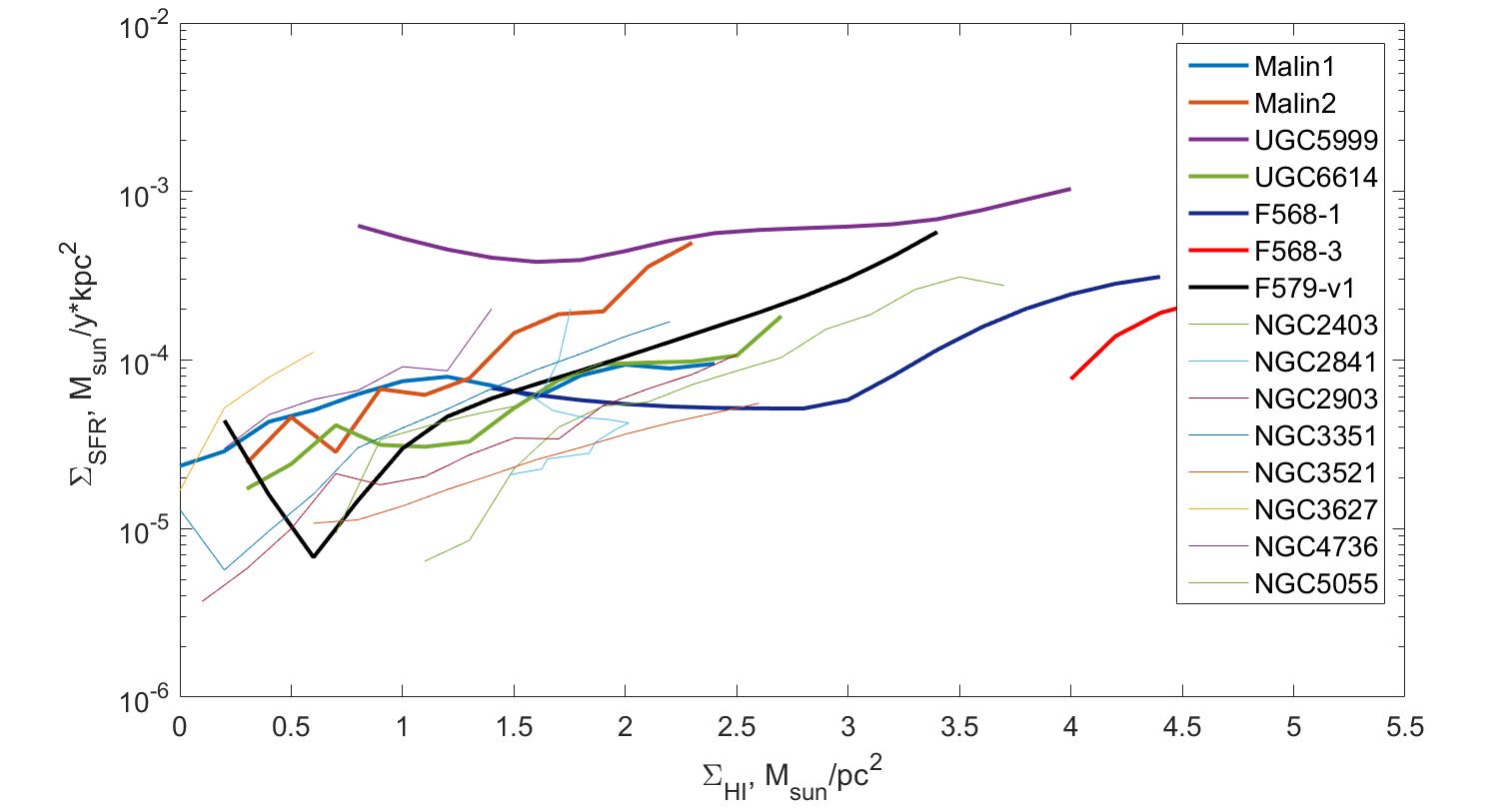}}
\caption{ A comparison of $\Sigma_{SFR}$ and $\Sigma_{HI}$. Thick solid lines represent the sub-sample of LSB-galaxies, while thin solid lines relate to the outer regions of normal spiral galaxies. $\Sigma_{SFR}$ were taken from Leroy et al. (2008) for the THINGS sub-sample and calculated from the GALEX FUV profiles presented in Wyder et al. (2009), following Eq. 4 from that paper.}
\label{fig:3}
\end{figure*}

For a more detailed comparison of LSB disks and low-density extended disks in HSB galaxies, their azimuthally averaged radial distributions of $\Sigma_{SFR}$ and $\Sigma_{HI}$ are compared in Fig.~\ref{fig:3} (HSB disks are marked by thin lines there). The diagram confirms that there is no sharp distinction between LSB and the low density disks of HSB galaxies. At the same time, some LSB galaxies (UGC 5999, Malin-1, Malin-2, F579-v1) have systematically higher star formation rates over a wide range of $\Sigma_{HI}$ than the outer regions of HSB galaxies with similar surface gas densities. Their higher rate of star is not related to either the enhanced optical surface brightness or the lower value of the stability parameter $Q_T$: the disks of LSB galaxies, like the outer regions of HSB galaxies, have $Q_T\sim 3-5$ in a wide range of radial distance. However, it may reflect the higher (by a factor of several) density of the disks of such LSB-galaxies than would be expected from their low surface brightness, so that a gravitational field of a disk compresses the gas layer more strongly, favoring star formation. 

The idea of underestimating the density of LSB disks from surface brightness measurements was first proposed by Fuchs (2003), see also discussion in Saburova (2012); Saburova $\&$ Zasov (2013).. The most significant evidence that the lowbrightness disks may be denser than suggested by conventional models of their stellar population is the presence of structural details such as bars and spiral arms in some LSB galaxies, where disk self-gravity should be essential, as well as the similarity of the stability parameter $Q_T$ over a wide range of R for the disks of LSB and HSB galaxies, despite the large difference in their surface brightness.

The underestimation of the mass of LSB-disks may be due to the bottom-heavy initial mass function of the stars that make up the bulk of the present-day disks, leading to a higher $M_*/L$ ratio, or it may be caused by the presence of a significant amount of high angular-momentum dark matter in the outer regions of the disks, coming from cosmological filaments, in addition to the conventional baryonic disk components.

\section{Summary}

At the diagram ``Star formation efficiency - volume (or surface) disk density'' LSB galaxies continue the sequence constructed for HSB galaxies. However, for the regions of the lowest disk densities the relationships are characterized by a large scatter, i.e. the SFE there is loosely dependent on the disk parameters there. In some LSB galaxies, the observed star formation is more intense than in the extended outer regions of normal spiral galaxies with similar surface HI densities, which may indicate that the disk density of LSB galaxies is higher than it is expected from their low surface brightness.

\section*{Funding} 
This project is supported by the RScF grant 23-12-00146.

\bibliographystyle{aa}
\bibliography{template}
Abramova O.V. and Zasov A.V., 2012, Astronomy Letters, 38, p. 755

Fuchs B., 2003, Astrophys. Space Sci., 284, p. 719

Leroy A.K., Walter F., Brinks E., et al., 2008, Astron. J., 136, p. 2782

Lutz K.A., Kilborn V.A., Koribalski B.S., et al., 2018, Mon. Not. R. Astron. Soc., 476, p. 3744

Ostriker E.C., McKee C.F., Leroy A.K., 2010, Astrophys. J., 721, p. 975

Saburova A.S., 2012, Astronomical and Astrophysical Transactions, 27, p. 251

Saburova A.S., Chilingarian I.V., Kasparova A.V., et al., 2019, Mon. Not. R. Astron. Soc., 489, p. 4669

Saburova A.S., Chilingarian I.V., Kasparova A.V., et al., 2021, Mon. Not. R. Astron. Soc., 503, p. 830

Saburova A.S. and Zasov A.V., 2013, Astronomische Nachrichten, 334, p. 785

Thilker D.A., Bianchi L., Meurer G., et al., 2007, Astrophys. J., Suppl. Ser., 173, p. 538

Wang J., Kauffmann G., J´ozsa G.I.G., et al., 2013, Mon. Not. R. Astron. Soc., 433, p. 270

Wyder T.K., Martin D.C., Barlow T.A., et al., 2009, Astrophys. J., 696, p. 1834

\end{document}